\documentclass[12pt]{article}

\def\bi{\begin{itemize}}
\def\ei{\end{itemize}}
\def\bea{\begin{eqnarray}} 
\def\eea{\end{eqnarray}}
\def\be{\begin{equation}}
\def\ee{\end{equation}}
\def\line{\hbox to \hsize}    
\def\frac #1#2{{#1\over #2}}

\def\eval #1#2#3{{\langle#1\vert#2\vert#3\rangle}} 

\def\1{\mbox{\bf I}}

\def\bm#1{\mbox{\boldmath$#1$}} 

\newenvironment{Quote}
{\begin{list}{}{%
\setlength{\leftmargin}{10 pt}
\setlength{\rightmargin}{\leftmargin}}
\item[]}
{\end{list}}

     {\refstepcounter{exercisenumber} \begin{Quote}\small{\sl
     Exercise~\thechapter.\theexercisenumber\/}:}
{\end{Quote}}

{\begin{Quote}\small{\sl Example\/}:}
{\end{Quote}}

\def\levelonelist{
        \begin{list}{\mybulA}%
                        {
        \setlength{\topsep}{0pt}
        \setlength{\parsep}{0pt}
        \setlength{\partopsep}{0pt}
        \setlength{\itemsep}{0pt}
                        }
                }
\def\leveltwolist{
        \begin{list}{\mybulB}%
                        {
        \setlength{\topsep}{0pt}
        \setlength{\parsep}{0pt}
        \setlength{\partopsep}{0pt}
        \setlength{\itemsep}{0pt}
                        }
                }

\def\el{\end{list}}

\usepackage{amsfonts}   
\usepackage{amssymb}
\usepackage{centernot}
\usepackage{microtype}
\usepackage{appendix}
\usepackage{hyperref}
\usepackage{bm}
\newcommand{\fsl}[1]{{\centernot{#1}}}

\begin{document}

\title{Gravitational Landau Levels  and the Chiral Anomaly}

\author{
Michael  Stone$^1$\\
\and
Porter Howland$^2$ \\ 
\and   
JiYoung Kim$^3$\\
\\
University of Illinois, Department of Physics\\ 1110 W. Green St.\\
Urbana, IL 61801 USA}   
\date{%
    $^1$m-stone5@illinois.edu\\%
    $^2$pbh2@illinois.edu \\%
     $^3$jkim623@illinois.edu\\[2ex]%
    \today
}


\maketitle

\begin{abstract} 
A popular  physical   picture of the mechanism   behind the  four-dimensional chiral anomaly is provided    by  the massless  Dirac equation in the presence of  constant electric and magnetic  background fields. The magnetic  field creates highly degenerate Landau levels, the lowest  of which is gapless. Any  parallel component of the electric field drives  a spectral flow in the gapless mode that causes particles to emerge from, or disappear into, the Dirac sea.  
Seeking   a similar  picture  for the gravitational contribution to the chiral anomaly, we    consider  the massless Dirac  equation in a background spacetime  that creates {\it gravitational\/}  Landau levels.  We find that in this case the resulting spectral flow, with its explicit particle production, accounts for  only a small part of the anomalous creation of chiral charge. The balance is provided by the vacuum expectation charge arising from the spectral asymmetry.   

\end{abstract}


 \section{Introduction}
\label{SEC:introduction}
When    charge-$e$  right-handed Weyl   field evolves  in  both a background  electromagnetic field $F_{\mu\nu}$ and a gravitational field  described by Riemann curvature tensor ${R^\alpha}_{\beta\mu\nu}$ the   associated chiral particle-number current $J^\mu_R$ is { anomalous\/}. Although classically we expect the   current to be conserved, the quantum operator has a non-zero divergence \cite{kimura,salam,eguchi} 
 \be
 \nabla_\mu J^\mu_R = -\frac{e^2} {32 \pi^2} \frac{\epsilon^{\mu\nu\rho\sigma}}{\sqrt -g}  F_{\mu\nu}F_{\rho\sigma}- \frac {1}{768\pi^2}\frac{\epsilon^{\mu\nu\rho\sigma}}{\sqrt -g}{R^\alpha}_{\beta\mu\nu} {R^{\beta}}_{\alpha\rho\sigma}.
 \label{EQ:anom-number}
 \ee
On the right-hand side of (\ref{EQ:anom-number}) both the   electromagnetic Chern-character   density $\propto {\epsilon^{\mu\nu\rho\sigma}} F_{\mu\nu}F_{\rho\sigma}$ and  the gravitational 
Pontryagin  density  $\propto {\epsilon^{\mu\nu\rho\sigma}}{R^\alpha}_{\beta\mu\nu}{R^{\beta}}_{\alpha\rho\sigma}$ serve as sources for  chiral charge    that  apparently appears from nowhere.

The physical mechanism behind the  term   involving the electromagnetic field  is well understood  \cite{kiskis,peskin}. In flat space with $(-,+,+,+)$ metric and $\epsilon^{0123}=+1$ we have 
\be
-\frac{e^2} {32 \pi^2}\epsilon^{\mu\nu\rho\sigma}  F_{\mu\nu}F_{\rho\sigma}= \frac{e^2}{4\pi^2}{\bf E}\cdot {\bf B}.
\ee
In the case of a uniform ${\bf B}$  the quantized cyclotron motion  in the plane perpendicular to ${\bf B}$ leads to  Landau levels with energy 
\be
\varepsilon (m,k_\parallel)= \pm \sqrt{2 eBm+k_\parallel^2}, \quad m\in {\mathbb N}_+,
\ee
Here $k_\parallel$ is the momentum component parallel to ${\bf B}$,
and each energy level has   degeneracy    $e|{\bf B}|/2\pi$  per unit area perpendicular to ${\bf B}$.   Considered as functions of $k_\parallel$ these $m>0$ levels   are   gapped, but there is   an additional    $m=0$ Landau level with the same degeneracy that  is distinguished  from the $m>0$ levels  in that it is both gapless and only one sign is allowed:  the one  with  $\epsilon = +k_{\parallel}$.  A  component  of the electric field parallel to ${\bf B}$  causes    $k_{\parallel}$  to increase  at a rate $\dot   k_{\parallel}=eE_{\parallel}$  so   the gapless modes   rise   through the $\varepsilon =0$ Fermi energy at a rate $eE_{\parallel}/2\pi$  per unit length. The net result is that right-handed particles are seen to emerge  out of the Dirac-sea vacuum   at a rate of   ${e^2}{\bf E}\cdot {\bf B}/{4\pi^2}$ per unit volume per unit time. Left-handed Weyl particles disappear  {\it into\/} the sea at the same rate. For Dirac particles the axial current 
\bea
J^\mu_5= J_R^\mu-J^\mu_L
\eea 
is therefore also anomalous  with coefficients that are twice those in  (\ref{EQ:anom-number}).   The  vector current $J^\mu= J^\mu_R+J^\mu_L$ is conserved.

There appears to be no comparably   simple  {physical\/} picture of  the gravitational   contribution to the anomaly arising from the Pontryagin term. The formal {mathematics\/}  is    understood {\it via\/} the  index theorem and Euclidean-signature  zero modes \cite{alvarez-gaume-ginsparg} but  the  necessity of    tidal forces encoded in the curvature tensor obscures the  physical mechanism. 
In particular  there  is no  {\it exact\/}    gravitational  analogue of the constant ${\bf E}$, ${\bf B}$ field configuration that we can use to aid  our understanding.

In the present  paper  we consider a family of space-times that come {\it close\/} to the ${\bf E}$, ${\bf B}$ case  in that they support a gravitational version of Landau levels and for which the rate of level crossing can be calculated and compared with that predicted by the anomaly equation.  We will find that despite the similarity of the Landau-level picture there are some surprising differences between the gauge-field and gravitational cases. 
 
In section  \ref{SEC:Heisenberg} we introduce the   space-time  geometry and its  analogue  Landau levels.  We then allow the metric to become time dependent  and compare the resulting chiral  {\it particle\/} creation with the total chiral {\it charge\/} creation  predicted by the anomaly. We find a substantial difference. In section \ref{SEC:index} we account for the  discrepancy by showing that  the vacuum   charge  arising from   the spectral asymmetry is much larger than the explicit particle creation.  

The  detailed calculations   are somewhat  intricate so we have relegated most of them to appendices.   Appendix \ref{SEC:appendix-heisenberg} explains the group theory behind the geometric and spectral  properties of the  spatial  part of our spacetime. Appendix \ref{SEC:appendix-dirac} provides a review of the Dirac equation in curved space-time, and constructs and solves the Weyl equation  for our family of spacetimes. Appendix \ref{SEC:eta-zero} computes the spectral asymmetry directly from the Weyl Hamiltonian eigenspectrum thus    demonstrating  the consistency of  the anomaly equation with  the total  chiral charge and also  identifying which eigenstates contribute the most to the vacuum charge.

 \section{Gravitational Landau levels }
\label{SEC:Heisenberg}

We seek a space-time geometry that has similar physics to the constant ${\bf E}$,  ${\bf B}$, gauge field configuration that gave us the physical picture of spectral flow creating particles out of the Dirac sea. Our    example  is based on      the metric  
\bea
ds^2 =- dt^2   +a(t)^2 (dx^2 +dy^2) + b(t)^2 (dz-xdy)^2
\label{EQ:heisenberg-metricI}
\eea 
 which  describes the geometry of a Bianchi type-II spacetime  associated with the Heisenberg group.
 
Consider first   the static case in which  $a$ and $b$ (both assumed positive) are independent of time. The  associated scalar field  Hamiltonian is  $H= -\nabla^2$ where $\nabla^2$ is 
the   Laplacian   
\bea
\nabla^2 
&=&  \frac 1{a^2} \left(\frac{\partial^2}{\partial x^2}+\left(\frac{\partial}{\partial y}+x\frac{\partial}{\partial z}\right)^2\right)+ \frac 1{b^2}\frac{ \partial^2}{\partial  z^2}.
\eea

To obtain a discrete set of eigenfunctions of $H$ we  impose  periodic boundary conditions 
\be   
 \psi(x,y,z)= \psi(x+l, y+m, z+n+ly),\quad l,m,n \in {\mathbb Z}
 \label{EQ:pbc}
\ee
on the wave-functions. 
The reason for this seeming odd choice is that each of  the $t={\rm constant}$ space slices described by (\ref{EQ:heisenberg-metricI})   can be identified with  the group manifold of   the Heisenberg Lie group. As such,   it possesses three  non-obvious translation symmetries arising from the group action. The periodicity (\ref{EQ:pbc}) is    the  simplest that  is compatible with these symmetries,  and is   the gravitational analogue of the twisted  periodic boundary conditions  required when we solve the Schr\"odinger equation in the presence of a Landau-gauge uniform magnetic field. The  construction of  mode expansions that automatically satisfy  these conditions is  described in section \ref{SUBSEC:harmonic}.  Note that while the $x,y,z$ coordinates are now restricted to the unit cube, the volume of the compactified space is $a^2b$, and this can be as large as desired.

In   section \ref{SUBSEC:scalar-laplacian} we show that these   boundary conditions lead to a complete set of eigenfunctions of  $-\nabla^2$ that are parameterized by three quantum numbers. The first, $n$, labels the momentum in the $z$ direction and can take any integer value; the second,  $k$, labels  a  Harmonic-oscillator eigenstate and  can be any non-negative integer; the third, $m$,  is an integer in the range $0,1,\ldots, |n|-1$. The corresponding eigenvalues are  \cite{gordon-wilson}
\be
\varepsilon_{n,k} = \frac 1{a^2} (2\pi |n|(2k+1)) + \frac {(2\pi n)^2}{b^2},\quad \hbox{$|n|$-fold degenerate}.
\ee
The $|n|$-fold degeneracy arises because the eigenvalues are independent of $m$. These eigenstates  arise  from essentially the same mathematics as  the usual Landau-gauge Landau levels, so it is natural to refer to them as  {\it Gravitational Landau levels}.  
In the magnetic  case, though, the degeneracy is the same for all  levels  and equal to   the number  $N=eB_zL_xL_y/2\pi$ of magnetic flux units passing through the $L_x$-by-$L_y$  torus     in the  $x$-$y$  plane.  In  the   gravitational  version   the analogue of  ``$BL_xL_y$'' is unity, but  the effective charge of a particle interacting with gravity depends on its energy-momentum, and, as a result, the  particle's charge $e$  is replaced by  the $z$-momentum  $2\pi n$. The  number of ``gravitational flux units" passing through the unit square in the $x$-$y$ plane  becomes    $n$, and this determines the degeneracy.

\subsection{Weyl Hamiltonian\/}
\label{SEC:dirac-spectrum}

 The Hamiltonian for  a two-component right-handed Weyl fermion moving in the metric   (\ref{EQ:heisenberg-metricI})  is  constructed in  section \ref{SUBSEC:spin-connection}, and is
\be
H_R=- i\left \{\frac 1 a\Big( \sigma_1 \partial_x +\sigma_2 (\partial_y +x \partial_z)\Big) + \frac 1b \sigma_3 \partial_z\right\}- \frac{b}{4a^2}{\mathbb I}_2.
\label{EQ:weyl-equation}
\ee
The diagonal term $ -({b}/{4a^2}){\mathbb I}_2$  arises from  the spin connection and can be interpreted as a geometry-dependent chemical potential. It will be the principal source of  spectral flow.
The eigenstates of (\ref{EQ:weyl-equation}) are obtained   in    section \ref{SUBSEC:dirac-spectrum}. Here   we  simply summarize the results.

\noindent The spectrum of $H_R$ falls into three broad classes   with two subclasses: 

 On states of the form 
\be
\chi=\left[\matrix{ \alpha \cr \beta}\right]e^{2\pi i( n_1x+n_2y)}, \quad n_1, n_2\in {\mathbb Z}.
\ee
we have
eigenvalues 
\be
\varepsilon(n_1,n_2)= -\frac{b}{4a^2}\pm \frac{2\pi}{a}\sqrt{n_1^2+n_2^2},
\ee
with one-state for each pair $(n_1,n_2)$.

On 2-spinors  of the form 
\be
\chi=\left[\matrix{ u(x) \cr v(x)}\right]e^{2\pi im y} e^{2\pi i nz}, \quad n\in {\mathbb Z}\setminus \{0\}, \quad m=0, \ldots |n|-1
\ee
there are two  different cases depending on whether the $z$-momentum  $n$ is positive or negative.

\noindent{Case i) ${ n> 0}$}: The  eigenvalues are
\be
\varepsilon_+(n,k) = -\frac{b}{4a^2}\pm \frac 1 a \sqrt{(2\pi n a/b)^2+ 4\pi |n| k}, \quad k\ge 1.
\ee
These  do not depend on $m$ so  each  level  has degeneracy $|n|$ and  the  corresponding eigenstates are the  Weyl-equation  version  of the gravitational  Landau levels.

The special case $k=0$ has   
\be
\chi_{0+}= \left[\matrix{ 0 \cr  \varphi_{0}}\right]e^{2\pi im y} e^{2\pi i nz},
\ee
where $\varphi_{0}$ is a harmonic-oscillator ground state wavefunction, 
and  eigenvalues
\be
\varepsilon(n,0) = - \frac{2\pi n}{b}-\frac{b}{4a^2}. \quad n>0, \quad \hbox{degeneracy $|n|$}.
\ee
From our experience with  the  magnetic field case, we might anticipate that these  will be ``topological'' gapless modes that provide the spectral flow.

\noindent { Case ii) ${ n<0}$}: We again have eigenvalues 
\be
\varepsilon_-(n,k) = -\frac{b}{4a^2}\pm \frac 1 a \sqrt{(2\pi n a/b)^2+ 4\pi |n| k}
\ee
that appear similar to the $n>0$  case, but the  $k=0$ eigenstates   are now  
\be
\chi_{0-}= \left[\matrix{ \varphi_{0}\cr0 }\right]e^{2\pi im y} e^{2\pi i nz}
\ee
with  eigenvalues 
\be
\epsilon(n,0) = \frac{2\pi n}{b}-\frac{b}{4a^2}, \quad n<0, \quad \hbox{degeneracy $|n|$}
\ee

Observe   that  ``topological" $k=0$ states have strictly negative energy eigenvalues 
 \be
 \epsilon(n)= -\frac{2\pi |n|}{b}-\frac{b}{4a^2}.
 \label{EQ:topological}
\ee
The reason for the $|n|$ in (\ref{EQ:topological}) is that changing the sign of the $z$-momentum changes the direction of the gravity-analogue of the magnetic flux.
 Being always negative, these energy levels   {\it never\/} cross  the  $\varepsilon=0$ Fermi level  --- so  our anticipation that the ``topological'' states provide the spectral flow was  premature. This is our first surprise.
 
 \subsection{Counting level-crossings }
 \label{SUBSEC:counting} 
 
 We have found a gravitational analogue the ${\bf B}$ in the $e^2{\bf E}\cdot {\bf B}/4\pi^2$ anomaly term.  To find an analogue of ${\bf E}$ we allow $a$ and $b$ to vary slowly    with time and seek the anomaly by  counting  the  number of states that cross zero energy and emerge from the Dirac sea.    There are two sources of Fermi-energy  level crossing: i) the ``chemical potential" $- b/4a^2$   will shift up or down,  ii) the individual eigenstate energies will shift because of the changing parameters in  the $\pm \sqrt{\ldots}$ expressions. 
 
 The symmetry  point at which the    $\pm \sqrt{\ldots}$ terms become zero   always lies below the $\varepsilon=0$ Fermi energy.  No levels  cross the symmetry point when  $a$, $b$ change, so the  number of states that cross  from the $\varepsilon<0$ Dirac  sea to positive energy  is   equal to the \underline{decrease}  in the number of states lying between the $\pm$ symmetry  point  and the $\varepsilon=0$ Fermi level.  
 Counting  the  exact number states in this energy-range is not easy, but if we assume that $b\gg a$ 
 we can use approximate density-of-states methods. 
 
 For the $n=0$, states with  
\bea
\varepsilon(n_1,n_2)&=& \frac{2\pi}{ a} \sqrt{n_1^2+n_2^2}- \frac{b}{4 a^2}\nonumber\\
&=& \frac{2\pi}{ a} \|{\bf n}\|- \frac{b}{4 a^2}
\eea
have $\varepsilon=0$ when 
\be
\frac{b}{4 a^2}= \frac{2\pi}{a}\|{\bf n}_{\rm max}\|.
\ee
So   
\be
\|{\bf n}_{\rm max}\|= \frac{1}{8\pi}\frac{b}{a},
\ee
and the   approximate  total  is
\be
N_{\rm periodic}\approx  \pi \|{\bf n}_{\rm max}\|^2= \frac{1}{64\pi} \left(\frac ba \right)^2.
\ee
 
 The Landau-level states with energy 
\be
\varepsilon(n,k)= \sqrt{\frac{4\pi^2 n^2}{b^2}+\frac{4\pi|n| k}{a^2}}- \frac b{4a^2}, \quad k\ge 1,
\ee
 have $\varepsilon(\pm |n|_{\rm max},k)=0$ at  
\be
 |n|_{\rm max}(k)=\frac{b^2}{2\pi a^2}\left(\sqrt{k^2+\frac{1}{16 }}-k\right).
\ee
So, taking into account the $|n|$-fold degeneracy, we have  
\bea
N_{\rm Landau}&=&\sum_{k=1}^\infty \sum_{n=-n_{\rm max}(k)}^{n_{\rm max}(k)} |n|\nonumber\\
&\approx&   \sum_{k=1}^\infty [n_{\rm max}(k)]^2\nonumber\\
&=& \frac 1 {4\pi^2  C}\left(\frac{b^4}{a^4}\right). 
\eea 
A numerical evaluation of the sum over $k$ gives $C\approx 635$.  

For   $b\gg a$ the number of level crossings from the $n=0$ states is negligible compared to those from the Landau-levels   because  $b^4/a^4$ is much larger than  $b^2/a^2$.   

Suppose now   we start at time $-\infty$ with a stationary  metric,   have a period of time in which $a$ and $b$ vary slowly, then  return to  a stationary  metric by $t=\infty$, then  the  estimated number of Landau-level level crossings is
 \be
N_{\rm crossing} \approx -\frac 1 {4\pi^2  C}\left[ \frac{b^4}{a^4} \right]_{-\infty }^{\infty}.
\ee
The minus sign appears  because an \underline{increase} in the number between $\varepsilon=0$ and the symmetry point means a \underline{decrease} in the number of levels above $\varepsilon=0$.

We can compare this result with that  expected from the  
 anomaly.  The gravitational source term in the anomaly is the four-form part,   $A_1$, of the total A-roof genus that that appears in the general Dirac index theorem  \cite{AS,getzler}.
 
 For general time-dependent $a$, $b$ we can use a  symbolic tensor analysis package such as {\tt ccgrg\/} \cite{ccgrg} to compute   
 \bea
\hat A_1&\stackrel{\rm def}{=}& \left(\frac{1}{768\pi^2} {R^{\alpha}}_{\beta\lambda\mu} {R^{\beta}}_{\alpha\nu\sigma}\right)dx^\lambda dx^\mu dx^\nu dx^\sigma\nonumber\\
&=& \frac{1}{48 \pi^2a^5}(a\dot b-b\dot a)\left(b^3 + 
 a^2 b (-\dot a^2 + a \ddot a)+a^3(\dot a\dot b-a\ddot b)\right)dtdxdydz. \nonumber\\
\eea
The coefficient of $dtdxdydz$ is an exact derivative
\bea&&\frac{1}{48 \pi^2a^5}(a\dot b-b\dot a)(b^3 + 
 a^2 b (-\dot a^2 + a \ddot a)+a^3(\dot a\dot b-a\ddot b)) \nonumber\\
 &=&\frac{d}{dt}\left( \frac{b^4-2a^2b^2 \dot a^2+4a a^3 b \dot a\dot b-2a^4 \dot b^2}{192 \pi^2 a^4}\right)
 \label{EQ:A-roof-integral}
 \eea
 as expected of a topological quantity.
 
 The coefficient  in the anomaly equation is \underline{minus} the A-roof expression because, as we will see in the next section,  the topological index  that counts zero modes has a plus sign for right-handed energy levels that sink into the Dirac sea  and a minus sign for  levels that emerge from the sea. If  we again start at time $-\infty$ with a stationary  metric,   have a period of time in which $a$ and $b$ vary slowly, then  return to  a stationary  metric by $t= \infty$,  then the anomaly equation tells us  change in the total charge 
 for   right-handed particles should be 
 \bea
 \Delta Q_R&=& \int_{-\infty}^{ \infty} dt \int_{[0,1]^3} dxdydz\left\{ \frac{-1}{48 \pi^2a^5}(a\dot b-b\dot a)\left(b^3 + 
 a^2 b (-\dot a^2 + a \ddot a)+a^3(\dot a\dot b-a\ddot b)\right)\right\}\nonumber\\  
 &=& -\frac1{192\pi^2}\left[\frac{b^4}{a^4}\right]_{-\infty}^{\infty}.
 \label{EQ:anomaly-prediction}
 \eea
 This is a much larger number than the estimated number of particles 
  \be
N_{\rm crossing} \approx -\frac 1 {4\pi^2  C}\left[ \frac{b^4}{a^4} \right]_{-\infty}^{\infty}
\ee
 created by the spectral flow.
This particle-creation shortfall is our second surprise.
 
 \section{Vacuum charge and the eta-invariant.}
 \label{SEC:index} 
 
 We need to account for the discrepancy between the chiral charge arising from the anomaly equation and the one arising from   explicit particle creation via spectral flow. We will see that  the discrepancy is explained by  the  vacuum-state  contribution to the  charge.
 
 We start with a rapid review  of the connection between the anomaly and the Dirac index theorem. 
When  the metric is of the form
\be
ds^2= d\tau^2+g_{ij}dx^idx^j  
\ee
the Euclidean-signature  Dirac operator appropriate for a  pair of left and right Weyl fermions is  (see \ref{SUBSEC:general-dirac})  
\be
{\fsl D}= 
\left[\matrix{0 & \partial_\tau +H\cr \partial_\tau -H&0} \right]
\ee
where, in flat space, 
\be
H= - i\sigma_a \partial_a = {\bm \sigma}\cdot {\bf p}
\ee
is the Hermitian Hamiltonian of a right-handed Weyl fermion. This remains true in curved space. Similarly $-H$ is the Hamiltonian of a left-handed Weyl fermion.

It is this  Euclidean-signature  Dirac operator that is related, {\it via\/} the Atiyah-Singer  index-theorem, to the spectral-flow interpretation of the anomaly. When the geometry changes slowly\footnote{A slow evolution of the energy levels allows us to avoid the full complexity of the  Lorentz-signature version of the index theorem \cite{baer}. This is safe for our purposes because   the  index  is a topological invariant.}, an instantaneous  Hamiltonian can still be defined but it now depends on $\tau$. When a  family of  eigenfunctions $u(\tau)$  of $H(\tau)$ have  eigenvalues $\varepsilon(\tau)$ that change sign from {positive} to {negative} as $\tau$ goes from $-\infty$ to $\infty$  the operator ${\fsl D}$  has a normalizable $\gamma^5\to +1$ zero mode\footnote{The ``$\sim$'' is because  by  ignoring the $\tau$ dependence of $u(\tau)$ we are making the slow-variation  adiabatic approximation.} 
\be
\Psi_{0,R}(\tau)\sim \left[\matrix{u\exp\{ \int_{0}^\tau \varepsilon(\tau')d\tau' \} \cr  0}\right].
\ee
Conversely, when a family of  eigenfunctions $v(\tau)$  of $H(\tau)$ have  eigenvalues $\varepsilon(\tau)$ that  change sign from {negative } to {positive } as $\tau$ goes from $-\infty$  to $\infty$ we find  a normalizable $\gamma^5\to -1$ zero mode 
\be
\Psi_{0,L}\sim \left[\matrix{0\cr v \exp\{  - \int_{0}^\tau \varepsilon(\tau')d\tau' \} }\right].
\ee
The {\it Dirac  index\/} is the difference $n_+-n_-$  in the number of normalizable right- and left-handed zero modes of ${\fsl D}$.  

For a closed Euclidean-signature manifold $M$ the index is given by 
\be 
 (n_+-n_-)= \int_M \hat A
 \label{EQ:AS-index}
 \ee
 where  the integral on the RHS  is necessarily a whole number.
Our manifold is  {\it not\/}  closed, but is of the form  $M=N\times {\mathbb R}$, where  ${\mathbb R}$ is  Euclidean time parameterized by $\tau$. In such an open manifold $\int_M \hat A$ is no longer guaranteed to be integral   but the index on the LHS of (\ref{EQ:AS-index}) must  still be an integer.  A generally non-integral {\it eta-invariant} was introduced by  Atiyah, Patodi and Singer  \cite{APS1} to  supply the difference.
The eta invariant is obtained from  
  \be
\eta_H(s) = \sum_{\lambda\ne 0} {\rm sgn}(\lambda)|\lambda|^{-s},
\ee
where the  sum is over the spectrum of $H$ (counted with multiplicity).  The sum is convergent for large  ${\rm Re}(s)$ and   has a meromorphic continuation to all of $s\in {\mathbb C}$. The 
eta-invariant  itself is defined by  $\eta_H= \lim_{s\to 0} \eta_H(s)$ and is always finite.

The APS theorem \cite{APS1} for  Dirac operators is now  the statement that 
\be
n_+-n_- = \int_{N\times(-\infty,\infty)} \hat A -\frac 12\left[{\rm dim\, Ker}(H)+\eta_H\right]_{\tau=-\infty}^{\tau=\infty}.
\ee
(The ${\rm dim\, Ker}(H)$ term is only present when an energy level is precisely zero, so we will  ignore it.)

After multiplication by two we  can rewrite the APS formula  as 
\be
-2\int_{N\times( -\infty,\infty)} \hat A= -2(n_+-n_-) -\left[\eta_H\right]_{\tau=-\infty}^{\tau=\infty},
\ee
in which the LHS is the integral over space-time of the RHS  of anomaly equation 
\be
 \nabla_\mu J^\mu_5 = - \frac {1}{384\pi^2}\frac{\epsilon^{\mu\nu\rho\sigma}}{\sqrt -g}{R^\alpha}_{\beta\mu\nu} {R^{\beta}}_{\alpha\rho\sigma}
 \label{EQ:anom-number2}
 \ee
for the chiral current of a \underline{Dirac} fermion which comprises  both a left- and a right-handed Weyl fermion.  If the anomaly was solely due to spectral flow the change in the total chiral charge would be 
\be 
\Delta Q_5= \Delta Q_R-\Delta Q_L\stackrel{?}{=} -2(n_+-n_-)
\ee 
because an eigenvalue of $H$  crossing from negative  to positive  means one extra  right-handed particle and one fewer  left-handed particle, and at the same time  adds unity to $n_-$. An eigenvalue of $H$  crossing from   positive to negative  means one fewer  right-handed particle and one more  left handed particle, and  adds unity to $n_+$.   

The $\eta$-invariant correction can therefore be interpreted  as the vacuum contributions 
\bea
Q_{R,\rm vacuum} &=&
 \int d({\rm vol})  \eval{0}{J^0_R}{0}_{\rm reg}
= -\frac 12  \eta_H\nonumber\\
Q_{L,\rm vacuum} &=& 
 \int  d({\rm vol})  \eval{0}{J^0_L}{0}_{\rm reg}= +\frac 12  \eta_H
\eea
that must be added to the count  of explicitly-created particle pairs in order to  compute the total chiral charge.  The interpretation of $-\eta(0)/2$ as a non-integral vacuum charge is of course familiar from the theory of charge fractionalization \cite{niemi-semenoff}. The surprise here is that $-\eta(0)/2$ is not a small fractional correction, but actually supplies the bulk of the anomaly-induced charge. We are tempted to    conjecture  that this charge is   due to the large spectral asymmetry from the ``topological'' modes that lie solely below the Fermi level.

 We can investigate the conjecture by observing that, when    $b/4a^2$ is  sufficiently small that the chemical potential lies so close   to the   $\pm \sqrt{\ldots}$  symmetry point  that  no level crossing has occurred, we should have 
 \be
 \int_{N\times(-\infty,\infty)} \hat A= \left[\frac 12 \eta_H(0)\right]_{\tau=-\infty}^{\tau=\infty}.
 \ee
 In  appendix \ref{SEC:eta-zero} we  compute $\eta_H(0)$ in the small-$b/4a^2$ region directly from the Weyl-Hamiltonian eigenvalue  spectrum 
and find that\footnote{Once  levels {\it have\/} crossed the LHS of (\ref{EQ:eta-repeat}) will become  
 $
2(n_+-n_-) + \eta_H(0)$. See \cite{gornet-richardson}. }
\be 
\eta_H(0)= \frac 1{96 \pi^2} \left(\frac{b^4}{a^4}\right)+ \frac 16.
 \label{EQ:eta-repeat}
 \ee
  which is consistent with (\ref{EQ:A-roof-integral}) and the induced charge from (\ref{EQ:anomaly-prediction}). We also   make the decomposition 
  \be
 \eta_H(0)=  \left(\frac{b^4}{a^4}\right) \frac 1{\pi^2} \left(\frac 1{48} + \frac 1{192}- \frac 1 {64}\right)
 +\frac 16
  \ee 
 of the eta invariant into parts due to the different branches of the spectrum. The   $-1/64$ is the part due to the always-negative $k=0$ modes, so, although not dominant, the do play a significant part in creating the anomalous chiral charge.

\section{Discussion}
\label{SEC:discussion}

We have constructed a family of space-times that possess gravitational analogues of Landau levels, and have used them to investigate the gravity contribution to the chiral anomaly.   Although there are similarities to the magnetic field case there are also some surprises. The first is that although  there are analogues of the topological modes that play a key role in the ${\bf E}\cdot {\bf B}$ electromagnetic case these modes do not contribute to any particle creation through  level crossing. The second surprise is that when we let the metric vary (slowly) with time  the number of explicitly created particles is small compared to the vacuum charge arising from the spectral asymmetry.  The ``topological" modes do, however, contribute a significant amount to this vacuum charge.

 It must  be noted, however, that in curved space, the decomposition of the charge into explicit particle number and a vacuum contribution is   frame dependent.  For  some  theories this can be understood  by initially defining operators such as the current or energy-momentum tensor to be normal ordered with respect to the positive and negative frequencies as seen by an observer moving along a world line of a  time-like Killing vector. Different Killing vectors lead to different normal-ordered operators, and so the  intitially-defined operators do not transform   as tensors.  To obtain physically meaningful tensor quantities  we must add $c$-number terms to the normal-ordered operators.

A classic example of this occurs in Hawking radiation  where a  Schwarzschild time-coordinate observer sees the  asymptotic energy  being carried by actual particles while in   Kruskal coordinates the   operator part of the energy momentum tensor has vanishing expectation value everywhere and the asymptotic energy flux comes entirely from the $c$-number vacuum  term --- see section III-A in \cite{stone-kim} for recent discussion of this. The  $-\eta(0)/2$ vacuum contribution to the chiral currents is an example of such a $c$-number.

\section{Acknowledgements} This work was not directly supported by any external funding agency, but it would not have been possible without resources provided by the Department of Physics at the University of Illinois at Urbana-Champaign.

\appendix
\appendixpage

\section{Heisenberg manifolds}
\label{SEC:appendix-heisenberg}

In this appendix we will review the  group theory that allows us to set-up and solve Dirac equation in the metric  (\ref{EQ:heisenberg-metricI}).  This subject has been extensively explored by mathematicians: see \cite{gordon-wilson}, and  in particular in \cite{ammann}.

\subsection{The Heisenberg Group}
The  three-dimensional Heisenberg Lie Group  ${\rm He}[{\mathbb R}]$ is the   set of  matrices   of the form
\be
g(x,y,z)=\left[\matrix{1& x& z\cr 0&1 &y\cr 0 &0& 1}\right], \quad x,y,z \in {\mathbb R}.
\ee
The entries $x,y,z$ can take any real values, so the  group manifold is the whole of ${\mathbb R}^3$.

The group product and inverse are given by 
\be  
\left[\matrix{1& x_1& z_1\cr 0&1 &y_1\cr 0 &0& 1}\right]\left[\matrix{1& x_2& z_2\cr 0&1 &y_2\cr 0 &0& 1}\right]= \left[\matrix{1& x_1+x_2& z_1+x_1 y_2+z_2 \cr 0&1 &y_1+y_2\cr 0 &0& 1}\right]
\ee
and 
\be
\left[\matrix{1& x& z\cr 0&1 &y\cr 0 &0& 1}\right]^{-1}= \left[\matrix{1& -x&xy- z\cr 0&1 &-y\cr 0 &0& 1}\right].
\ee

The associated Lie  algebra $\mathfrak {he}$ has elements
\be
\left[\matrix{0& \xi & \zeta \cr 0&0 &\eta \cr 0 &0& 0}\right]= \xi \hat X+\eta \hat Y+\zeta \hat Z
\ee
where the $\hat X, \hat Y, \hat Z $ matrices have commutators 
\be
[\hat X,\hat Y]=\hat Z,\quad [\hat X,\hat Z]= [\hat Y,\hat Z]=0.
\ee
The physicists'  Heisenberg algebra identifies  $\hat X\mapsto \hat q$, $\hat Y\mapsto \hat p$ and $\hat Z\mapsto i\hbar$ so $[q,p]=i\hbar$. 

The exponential map ${\rm Exp}: \mathfrak{ he}\to {\rm He}$ given by 
\be
{\rm exp}[\xi \hat X+\eta \hat Y+\zeta \hat Z]= \left[\matrix{1& \xi& \zeta + {\textstyle \frac 12}\xi \eta \cr 0&1 &\eta \cr 0 &0& 1}\right]
\ee
is a bijection and the  exponential coordinates and the original defining coordinates are  globally related by
\bea
x&=& \xi,\nonumber\\
y&=& \eta,\nonumber\\
z&=& \zeta+ \textstyle \frac 12 \xi\eta.
\eea

As with any Lie group, 
multiplication of a  point in the group manifold by   infinitesimal group elements  from the {\it right\/} gives  \underline{left}-invariant vector fields, which in this case are 
\bea
X&=& \partial_x, \nonumber\\
Y&=& x\partial_z+\partial_y,\nonumber\\
Z&=&\partial_z.
\eea
Their vector-field Lie brackets  obey  $[X,Y]=Z$ {\it etc\/}.

The corresponding left-invariant Maurer-Cartan forms are extracted from  
 \be
 g^{-1}dg=  \left[\matrix{1& -x&xy- z\cr 0&1 &-y\cr 0 &0& 1}\right]\left[\matrix{0& dx& dz\cr 0& 0&dy\cr 0 &0& 0}\right]=\left[\matrix{0& dx&dz -xdy \cr 0&0 &dy\cr 0 &0& 0}\right]
 \ee
 as 
 \bea
 \omega_X&=&dx,\nonumber\\
 \omega_Y&=& dy,\nonumber\\
 \omega_Z&=& dz-xdy.\nonumber
 \eea
 They satisfy  $\omega_X(X)=1$ and $\omega_X(Y)=0$ {\it etc.\/}, so  these are indeed the 1-forms dual to the  left-invariant vector fields. 
 
 Any constant-coefficient quadratic expression    in the left-invariant Maurer-Cartan forms   can serve  as a left-invariant metric, meaning that the Lie derivative ${\mathcal L}_X g=0$ {\it etc.\/},  so $X$, $Y$, $Z$ are Killing vectors.

Our metric  (\ref{EQ:heisenberg-metricI})  can be written as 
\bea
ds^2&=& a^2 \omega_X\otimes \omega_X+ a^2 \omega_Y\otimes \omega_Y+ b^2 \omega_Z\otimes \omega_Z,\nonumber\\
&=& a^2 dx^2 +a^2 dy^2 + b^2 (dz-xdy)^2,
\label{EQ:He-metric}
\eea
and is therefore left-invariant.

In terms of the exponential coordinates $x= \xi$, $y= \eta$ $z= \zeta+ \xi\eta/2$
this metric can be made to look   more complicated \cite{baerI}
\be 
ds^2= \left(a^2 + \frac {b^2\eta^2}{4}\right) d\xi^2+   \left(a^2 + \frac {b^2\xi^2}{4}\right)d\eta^2 + b^2 d\zeta^2
+ b^2\left(\eta d\zeta d\xi- \xi d\zeta d\eta - \frac { \eta\xi}{2} d\eta d\xi\right), 
\label{EQ:He-metric-exponential}
\ee
but of course it is just a reparametrization of  same space.

 In order to obtain   a countable set of normalizable  Dirac eigenfunctions it is useful  to compactify ${\mathbb R}^3$ by introducing periodic boundary conditions. 
 We   can do this by quotienting   by the left action of a  discrete subgroup of ${\rm He}[\mathbb R]$. We must use   a {\it left action\/}  because we want  our left-invariant metric to descend to a well-defined  metric on the quotient space.
  
The simplest discrete   subgroup of ${\rm He}[{\mathbb R}]$ is $\Gamma={\rm He}[{\mathbb Z}]$ with elements 
\be
\Gamma= \left[\matrix{1& l& n\cr 0&1 &m\cr 0 &0& 1}\right], \quad l,m,n \in {\mathbb Z}.
\ee
The right coset $\Gamma\backslash {\rm He}$  identifies 
\be
\left[\matrix{1& x& z\cr 0&1 &y\cr 0 &0& 1}\right]\sim  \left[\matrix{1& l& n\cr 0&1 &m\cr 0 &0& 1}\right]
\left[\matrix{1& x& z\cr 0&1 &y\cr 0 &0& 1}\right]=\left[\matrix{1& x+l& z+n+ly\cr 0&1 &y+m\cr 0 &0& 1}\right],
\ee
so functions on $\Gamma\backslash {\rm He}$ are functions  on ${\mathbb R}^3$ that satisfy  
\be
 f(x,y,z)= f(x+l, y+m, z+n+ly),\quad l,m,n \in {\mathbb Z}.
 \label{EQ:periodicBCs}
 \ee
The appearance of $y$ in the $z$ entry on the RHS of (\ref{EQ:periodicBCs}) means  that these are ``twisted'' periodic boundary conditions.  

If $\{x\}$ denotes the fractional part of $x$ and $\lfloor x\rfloor$ the integer part, we  can take $l= -\lfloor x\rfloor$ and $m= - \lfloor y\rfloor$, $n=- \lfloor z- \lfloor x\rfloor y\rfloor$ and  so replace any  group element by an  equivalent 
\be
\left[\matrix{1& x& z\cr 0&1 &y\cr 0 &0& 1}\right]\sim  \left[\matrix{1& \{x\}& \{z-\lfloor x \rfloor y \} \cr 0&1 &\{y\}\cr 0 &0& 1}\right].\
\ee
in which  the new $x,y,z$  each lie  in $[0,1)$. Consequently  the  compactified coset  manifold can be identified with    the cube  $[0,1]^3\in {\mathbb R}^3$, but  with twisted   glueings of the opposite faces.


\subsection{Harmonic analysis}
\label{SUBSEC:harmonic}

  The volume form for both systems of coordinates is 
\be
 {\rm Vol}=(a^2b) \omega_X\wedge \omega_Y \wedge \omega_Z= (a^2 b) dx\wedge dy \wedge dz =(a^2 b)d\xi  \wedge d\eta\wedge d\zeta
\ee
and provides  the   measure defining  a Hilbert space $L^2[\Gamma\backslash {\rm He}]$.

Our  twisted boundary conditions appear  to hinder Fourier transforms and mode decompositions on $L^2[\Gamma\backslash {\rm He}]$, but in fact they are no more complicated than  --- and indeed closely related to ---  the mode decompositions that are  used when solving for the eigenstates of a Schr{\"o}dinger particle moving in a Landau-gauge uniform magnetic field \cite{goldbart-stone}.    

What happens is that  the Hilbert space $L^2[\Gamma\backslash {\rm He}]$  decomposes into an orthogonal direct sum 
 \be
 L^2[\Gamma\backslash {\rm He}]= \bigoplus_{n\in \mathbb Z} {\mathcal H}_n
 \ee
 where 
 \be
 f\in {\mathcal H}_n \Leftrightarrow f(x,y,z+s)= e^{2\pi i ns} f(x,y,z).
 \ee
 Consequently ${\mathcal H}_n$ consists of  functions with discrete  $z$-momenta labelled by $n$.
 In particular, ${\mathcal H}_0$   consists of  ordinary periodic functions of $x$ and $y$ that do not depend on $z$.  
 
A  function    $f\in {\mathcal H}_1$  can be obtained from any  function $\psi(x)\in L^2[{\mathbb R}]$ by a Zak (or Weil-Brezin) transform \cite{Zak,weil-brezin}
 \be
 f(x,y,z)= W_1[\psi](x,y,z)\stackrel{\rm def}{=}  \sum_{L\in {\mathbb Z}} \psi(x+L) e^{2\pi i Ly} e^{2\pi i z}.
  \ee
  To see that this is so, observe that 
  \bea
  f(x+l, y+m, z+n+ly)&=& \sum_{L\in {\mathbb Z}} \psi(x+l+L) e^{2\pi i L(y+m)} e^{2\pi i (z+n+ly)}\nonumber\\
  &=& \sum_{L'\in {\mathbb Z}} \psi(x+L') e^{2\pi i (L'-l) y} e^{2\pi i (z+ly)}\nonumber\\
  &=& \sum_{L'\in {\mathbb Z}} \psi(x+L') e^{2\pi i L' y} e^{2\pi i z}\nonumber\\
  &=& f(x,y,z).
  \eea
  Indeed all functions in ${\mathcal H}_1$ can be obtained this way because given an element of $f\in {\mathcal H}_1$ we can obtain a corresponding   $\psi(x)$  via  an inverse map $W^{-1}_1:{\mathcal H}_1\to L^2[{\mathbb R}]$ given by 
  \be
  [W_1^{-1}f](x)= \int_0^1 f(x,y,0) dy\stackrel{\rm def}{=} \psi(x).
  \ee
  
  For $n\ne 0$,  ${\mathcal H}_n$  can be further decomposed as an orthogonal  direct sum
  \be
  {\mathcal H}_n =  \bigoplus_{m=0}^{|n|-1} {\mathcal H}_{n,m}.
  \ee
 Functions  in ${\mathcal H}_{n,m}$ obey
  \be
  f(x,y+ \frac{1}{n},z)= e^{2\pi i m/n} f(x,y,z).
  \ee
  and  may be   obtained from a $\psi(x) \in L^2[{\mathbb R}]$ as
  \be
   f(x,y,z)= W_{n,m}[\psi](x,y,z)\equiv  \sum_{L\in {\mathbb Z}} \psi\left(x+{L}+ \frac{m}{n}\right) e^{2\pi i (nL+m)y} e^{2\pi i n z}, \quad m=1,\ldots, n-1.
   \ee
  There is again an inverse map  $W^{-1}_{n,m}:{\mathcal H}_{n,m}\to L^2[{\mathbb R}]$  
  \be
   [W^{-1}_{n,m}f](x) = \int_0^1 e^{-imy} f(x-m/n,y,0) dy=\psi(x),
  \ee
  and so again all functions in ${\mathcal H}_{n,m}$ arise in this manner. 

\subsection{Scalar field Laplacian}
\label{SUBSEC:scalar-laplacian}
  
The ${\mathcal H}_n$ and ${\mathcal H}_{n,m}$  subspaces  correspond to the eigenspace decompositions of the Laplacian 
\bea
\nabla^2 &=& \frac 1{a^2} \left(X^2+Y^2\right)+ \frac 1{b^2} Z^2\nonumber\\
&=&  \frac 1{a^2} \left(\frac{\partial^2}{\partial x^2}+\left(\frac{\partial}{\partial y}+x\frac{\partial}{\partial z}\right)^2\right)+ \frac 1{b^2}\frac{ \partial^2}{\partial  z^2}.
\eea
In particular, 
when $-\nabla^2$ acts on functions in $f=W_{n,m}[\psi](x,y,z)\in {\mathcal H}_{n,m}$ we have
\be
-\nabla^2 f= \frac 1{a^2}\sum_{L\in {\mathbb Z}} \left(-\psi''\left(x+L+\frac mn\right)+(2\pi n)^2\left(x+L+\frac mn\right)^2 \psi\left(x+L+\frac mn\right)\right)+ \frac{(2\pi n)^2}{b^2} f
\ee
We recognize that the expression in parentheses is an $\omega_n = 2\pi |n|$ harmonic oscillator whose  eigenfunctions are $\psi_{n,k}(x+L+m/n)$ with
\be
\psi_{n,k}(x)=|\omega_n|^{1/4}\varphi_k(\sqrt{\omega_n} x).
\ee
Here
\be
\varphi_{k}(x)\equiv  \frac{1}{\sqrt{2^k k! \sqrt{\pi}}} H_k(x) e^{-x^2/2},
\ee
are the normalized $\omega=1$  wavefunctions.

The resulting  eigenvalues of $-\nabla^2$ are \cite{gordon-wilson}
\be
\mu_{n,k} = \frac 1{a^2} (2\pi |n|(2k+1)) + \frac {(2\pi n)^2}{b^2},\quad k\in {\mathbb N},\quad  \hbox{$n$-fold degenerate}.
\ee
The $n$-fold degeneracy arises because the eigenvalues are independent of $m=0,\ldots, |n|-1$.

 For functions in ${\mathcal H}_0$ we have the usual 2-d periodic spectrum
 \be
 \lambda_{m,n}= \frac {(2\pi )^2(k_x^2+k_y^2)}{a^2}, \quad k_x,k_y\in {\mathbb Z}.
 \ee

\section{Dirac equation in curved space-time}
\label{SEC:appendix-dirac}

\subsection{General manifolds}
\label{SUBSEC:general-dirac}
The skew-Hermitian Dirac  operator on an $N$-dimensional Euclidean-signature    Riemann manifold $(M,g)$ is  \cite{fock}
\be
{\fsl D}=  \gamma^a e^\mu_a \left(\partial_\mu + \textstyle{\frac 12} \sigma^{bc}\,\omega_{bc\mu}\right)= \gamma^a D_a. 
\ee
Here the 
${\bf e}_a\equiv e_a^\mu\partial_\mu$ compose  an orthonormal vielbein on   $M$ and 
the  $\gamma^a$ are Hermitian matrices obeying 
\be
\{\gamma^a,\gamma^b\}= 2\delta^{ab}.
\ee
The object  
\be
D_a \equiv  D_{{\bf e}_a}=e^\mu_a \left(\partial_\mu + \textstyle{\frac 12} \sigma^{bc}\,{ \omega_{bc}}_\mu\right)
 \ee
 is the covariant derivative acting on the components of a Dirac spinor. It contains 
 the ``spin-connection''  components $\omega_{bca}\equiv { \omega_{bc}}_\mu e^{\mu}_a $   defined by the action of the usual covariant derivative  on the vielbein vectors
 \be
e^\mu_a\nabla_\mu {\bf e}_c =  \nabla_{{\bf e}_a}{\bf e}_c=   {\bf e}^b{ \omega_{bca}}
 \ee
 combined with  the skew-Hermitian  spin-representation  generators of $\mathfrak {so}(N)$
\be
\sigma^{ab} = \textstyle{\frac 14} [\gamma^a,\gamma^b]
\ee
 that   obey
\bea
[\sigma^{ab},\sigma^{cd}]&=& \delta^{bc} \sigma^{ad}- \delta^{ac} \sigma^{bd}+\delta^{ad} \sigma^{bc}-\delta^{bd} \sigma^{ac},\nonumber\\
 {}[\sigma^{ab},\gamma^c]&=& \gamma^a \delta^{cb}- \gamma^b \delta^{ac}.
\eea

In four space-time dimensions  we can  take the {euclidean}-signature   gamma matrices to be 
\be
\gamma^5= \left[\matrix{1 &0 \cr 0 &-1}\right], \quad \gamma^0= \left[\matrix{0 &1 \cr 1 &0}\right].\quad
\gamma^a = \left[\matrix{0 &-i\sigma_a \cr i\sigma_a &0}\right],
\ee
The space parts of the spin operators are then
\be
\sigma^{ab}=\frac 14 [\gamma^a,\gamma^b]= \frac 12  \left[\matrix{\sigma_a\sigma_b &0 \cr 0 &\sigma_a\sigma_b}\right]= \frac i 2 \epsilon_{abc} \sigma_c \otimes {\mathbb I}_2, \quad a,b=1,2,3.
\ee

When the metric is of the form
\be
ds^2= d\tau^2+g_{ij}dx^idx^j  
\ee
we have
\be
{\fsl D}= 
\left[\matrix{0 & \partial_\tau +H\cr \partial_\tau -H&0} \right]
\ee
where, in flat space, 
\be
H= - i\sigma_a \partial_a = {\bm \sigma}\cdot {\bf p}
\ee
is the Hermitian Hamiltonian of a right-handed Weyl fermion. This remains true in curved space. Similarly $-H$ is the Hamiltonian of a left-handed Weyl fermion.

In  Minkowski-metric signature $(-,+,+,+)$  the gamma matrices  become   
\be
\gamma^5= \left[\matrix{1 &0 \cr 0 &-1}\right], \quad \gamma^0= \left[\matrix{0 &i \cr i &0}\right],\quad
\gamma^a = \left[\matrix{0 &-i\sigma_a \cr i\sigma_a &0}\right].
\ee
Then 
\be
{\fsl D} \left[\matrix{\psi_R \cr  \psi_L}\right]=\left[\matrix{0 & i\partial_t +H\cr i\partial_t -H}\right]\left[\matrix{\psi_R \cr  \psi_L}\right]= 0
\ee
becomes the pair of Weyl equations 
\bea
i\partial_t \psi_R = \phantom -H \psi_R\nonumber\\
i\partial_t \psi_L = -H \psi_L
\eea
The $H$ operator is the same in both Euclidean and Minkowski signatures.

\subsection{Weyl Hamiltonian    for the Heisenberg manifold}
\label{SUBSEC:spin-connection}

Define the normalized dreibeins $E_1= X/a$, $E_2= Y/a$, $E_3= Z/b$ with the commutators 
\be
[E_1,E_2]= \frac{b}{a^2} E_3, etc.
\ee
To find the spin-connection we can use the 
Koszul formula
 which extracts  the   Levi-Civita spin connection from
 \be
  2g(\nabla_X Y,Z)= Xg(Y,Z)+Yg(Z,X)-Zg(X,Y)-g(X,[Y,Z])+g(Y,[Z,X])+g(Z,[X,Y])
  \ee
  For our orthonormal frame the first three terms are zero, so the dreibein spin-connection  components  
  \be
  \omega_{ijk}= g(\nabla_{E_k} E_j,E_i)=\omega_{ij\mu}E_k^\mu
  \ee
 require only knowing the commutators  appearing in 
  \be
  \omega_{ijk}= {\textstyle \frac 12}\left\{-g(E_j,[E_k,E_i]) - g( E_k,[E_j, E_i],)+g(E_i,[E_k. E_j])\right\}.
  \ee
  The only non-zero components  are 
  \bea 
  \omega_{231} &=& - \frac{b}{2a^2}\nonumber\\
  \omega_{312}&=& - \frac{b}{2a^2}\nonumber\\
  \omega_{123}&=&+\frac{b}{2a^2}
  \eea 
  together with their $\omega_{ijk}= -\omega_{jik}$ partners.

We reviewed the curved space Dirac operator in section \ref{SUBSEC:general-dirac}.
For the three dimensional case we   set  $e^\mu_a\partial_\mu= E_a$ and $\gamma^a= \sigma_a$ so
\bea
\sigma^{12}&=& \frac 14[\sigma^1,\sigma^2]= \frac i2 \sigma_3\nonumber\\
\sigma^{23}&=& \frac 14[\sigma^2,\sigma^3]= \frac i2 \sigma_1\nonumber\\
\sigma^{31}&=& \frac 14[\sigma^3,\sigma^1]= \frac i2 \sigma_2
\eea
 giving 
\bea
{\fsl D}&=&\sigma_1(E_1+{\textstyle \frac 12} \sigma^{ab}\omega_{ab1}) + \sigma_2(E_2+{\textstyle \frac 12} \sigma^{ab}\omega_{ab2})+\sigma_3(E_3+{\textstyle \frac 12} \sigma^{ab}\omega_{ab3})\nonumber\\
&=& \sigma_1\left(E_1-\frac {ib} {4a^2} \sigma_1 \right) 
+ \sigma_2\left(E_2 - \frac {ib} {4a^2} \sigma_2\right)+\sigma_3\left(E_3+ \frac {ib} {4a^2} \sigma_3\right)\nonumber\\
&=&
\sigma_1 E_1+\sigma_2 E_2+\sigma_3 E_3- \frac{ib}{4a^2}.
\eea
When we extend to 4 dimensions $-i{\fsl D}$ will become the Hamiltonian of the right-handed Weyl fermion.
\be
H_R=- i( \sigma_1 E_1+\sigma_2 E_2+\sigma_3 E_3)- \frac{b}{4a^2}.
\ee
Similar geometric methods can be applied to derive Dirac operators and Hamiltonians on spacetimes  related to group manifolds. such as the   Bianchi type-IX spaces   of  deformed 3-spheres \cite{hitchin,gibbons} and their higher-dimensional Berger sphere versions.

\subsection{Weyl  spectrum for the Heisenberg manifold.}
\label{SUBSEC:dirac-spectrum}

In principle the two-component spinors on which $H_R$ acts could have non-trivial spin structures --- meaning that they  acquire additional factors of $\pm 1$ as they pass though the periodic boundary conditions \cite{ammann}. We will consider only the simplest case  in which both upper and lower components have the same periodicity properties as the scalar functions in the Laplace eigenproblem.
    
Acting on states of the form 
\be
\chi=\left[\matrix{ \alpha \cr \beta}\right]e^{2\pi i( n_1x+n_2y)}
\ee
we have
\be
H_R\to \frac {2\pi} a \left[ \matrix{0&  n_1-in_2 \cr n_1+in_2& 0}\right]- \frac {b}{4a^2} 
\ee
and hence eigenvalues 
\be
E(n_1,n_2)= -\frac{b}{4a^2}\pm \frac{2\pi}{a}\sqrt{n_1^2+n_2^2}.
\ee

On states of the form 
\be
\chi=\left[\matrix{ u(x) \cr v(x)}\right]e^{2\pi im y} e^{2\pi i nz}
\ee
we have
\be
H_R\to \frac 1 a \left[ \matrix{2\pi n a/b&  -i (\partial_x +2\pi n (x+m/n))\cr -i(\partial_x - 2\pi n(x+m/n))& - 2\pi na /b}\right]- \frac {b}{4a^2} 
\ee
Here there are two  different cases to consider:

\noindent{\bf Case ${\bf n> 0}$}: 

 Set $\omega_n = 2\pi |n|>0$ then on oscillator states $\varphi_k(\sqrt{\omega_n}(x+m/n))$ we recognize the ladder operators 
 \bea
 \partial_x+\omega_n x&=&\sqrt{2\omega_n} \hat a,\nonumber\\
 -\partial_x+\omega_n x &=& \sqrt{2\omega_n}\hat a^\dagger
 \eea
 making 
 \bea
( \partial_x+\omega_n x)\varphi_k&=& \sqrt{2\omega_n k} \varphi_{k-1}\nonumber\\
(\partial_x -\omega_n x)\varphi_{k-1}&=&-\sqrt{2\omega_n k} \varphi_{k}
\eea
We therefore try spinors of the form
 \be
 \chi= \left[\matrix{ \alpha \varphi_{k-1} \cr \beta \varphi_{k}}\right]e^{2\pi im y} e^{2\pi i nz}
\ee
whence
\be
H_R\to \frac 1 a \left[ \matrix{2\pi n a/b&  -i  \sqrt{2\omega_n k}\cr +i \sqrt{2\omega_n k}& - 2\pi na/b}\right]- \frac {b}{4a^2},
\ee
with eigenvalues 
\be
\mu_+(n,k) = -\frac{b}{4a^2}\pm \frac 1 a \sqrt{(2\pi n a/b)^2+ 4\pi |n| k}.
\ee
Again we have a degeneracy $m= 1,\ldots, |n|-1$ indicating the existence of Dirac gravitational Landau levels.

The special case $k=0$ has $\varphi_{k-1}=0$ and hence eigenvectors
\be
\chi_{0+}= \left[\matrix{ 0 \cr  \varphi_{0}}\right]e^{2\pi im y} e^{2\pi i nz}
\ee
with eigenvalues
\be
\mu_+(n,0) = - \frac{2\pi n}{b}-\frac{b}{4a^2}. \quad n>0.
\ee

\noindent {\bf Case ${\bf n<0}$}:

 Again  $\omega = 2\pi |n|$, but $\hat a$ and $\hat a^\dagger$ are interchanged so  we need 
\be
 \chi= \left[\matrix{ \alpha \varphi_k \cr \beta \varphi_{k-1}}\right]e^{2\pi im y} e^{2\pi i nz}
\ee
whence 
\be
H_R = \to \frac 1 a \left[ \matrix{2\pi n a/b&  +i  \sqrt{2\omega k}\cr -i \sqrt{2\omega k}& - 2\pi na/b}\right]- \frac {b}{4a^2}
\ee
with eigenvalues 
\be
\mu_-(n,k) = -\frac{b}{4a^2}\pm \frac 1 a \sqrt{(2\pi n a/b)^2+ 4\pi |n| k}
\ee
that appear the same as before, but the  special $k=0$ case 
\be
\chi_{0-}= \left[\matrix{ \varphi_{0}\cr0 }\right]e^{2\pi im y} e^{2\pi i nz}
\ee
now has  eigenvalues 
\be
\mu_-(n,0) = \frac{2\pi n}{b}-\frac{b}{4a^2}, \quad n<0
\ee

Note that for $n>0$ the special eigenvalues are $- \frac{2\pi n}{b}-\frac{b}{4a^2}$ and for $n<0$  they are $ \frac{2\pi n}{b}-\frac{b}{4a^2}$, so the general case is $ -\frac{2\pi |n|}{b}-\frac{b}{4a^2}$.
The $|n|$ agrees  with the spectrum  in \cite{ammann}
 but does  not accord with the spectrum  displayed on  page 30 of \cite{gornet-richardson}, which has $n$ rather than $|n|$.
 
 \section{Computing $\eta(0)$ from the spectrum}
 \label{SEC:eta-zero}
 
  We now     compute the spectral asymmetry $\eta(0)$ directly from the spectrum of the Weyl Hamiltonian\footnote{For recent discussion of other such calculations see \cite{dowker}.}. 
  We do this partly   to  compare it  with  the integral of the $\hat A$ genus, but more importantly so that  we can see how much of the vacuum charge arises  from the $k=0$ ``topological'' modes,  the  $k>0$ Landau-levels, and the $n=0$ modes. 
  Each of these  three parts of the spectrum requires different techniques,

 The simplest is the $k=0$ ``topological" branch with energies 
 \be
 \varepsilon = - \frac{b^2}{4a^2} - \frac {2\pi |n|}{b}, \quad  \hbox{degeneracy $2|n|$}
 \ee
 For this  we can make use of the   Hurwitz zeta function defined by 
 \be
 \zeta(s,x)= \sum_{n=0}^\infty  (n+x)^{-s}, \quad {\rm Re}(s)>1.
 \ee
 Its  analytic continuation to negative integers is  
 \be
 \zeta(-k, x)= - \frac{B_{k+1}(x)}{k+1}, 
 \quad k=0,1,2,\ldots
 \ee 
 where $B_{k}(x)$ are the Bernoulli polynomials. 
 Using this formula   we find  
 \bea
 \eta_{\rm top}(0) &\stackrel{\rm def}{=}&\lim_{s\to 0} \left\{-\sum_{n=0}^\infty  2n \left|  \frac{b}{4a^2} + \frac {2\pi |n|}{b}\right|^{-s}\right\}\nonumber\\
 &=& -2\left\{ \zeta(-1, b^2/8\pi a^2) - (b^2/8\pi a^2)\zeta(0, b^2/8\pi a^2)\right\}\nonumber\\
 &=&\frac 16- \frac 1{64 \pi^2} \frac{b^4}{a^4}.
 \label{EQ:eta-topological}
\eea

For the other branches with their $\pm \sqrt{\ldots}$ we use  the method described on page 34 of    \cite{hitchin}.   The   idea is  to assume that $\lambda=b/4a^2$ is small enough that  \bea
 -\lambda + \sqrt{\ldots}&>&0,\nonumber\\
 -\lambda - \sqrt{\ldots}&<&0,
 \eea
 ensuring that  there has been no level crossing.  Then,  with $E=+ \sqrt{\ldots}$,
 we define  auxiliary functions 
\be
F(s)= \sum_{E} ({\rm degeneracy}) |E|^{-s},
\ee
and expand $|E- \lambda|^{-s}- |E+\lambda|^{-s}$   as a  binomial series to find 
\be
\eta(s)=2s F(s+1) \lambda + \frac 13 s(s+1)(s+2) F(s+3)\lambda^3 +\ldots .
\label{EQ:hitchin-trick}
\ee
For $\lambda$ sufficiently small  the remainder denoted by ``$\ldots$"  will be analytic  and zero at  $s=0$.
Indeed the factors of $s$ ensure that the $s\to 0$ limit of all terms will be zero unless $F(s+1)$ and $F(s+3)$ have simple poles at $s=0$.   We must therefore compute the    residues of $F(s)$ at $s=1$ and $s=3$.  

For the periodic spectrum involving 
\be
E(n_1,n_2)=\left( \frac{2\pi}{a}\right) \sqrt{n_1^2+n_2^2}, \quad n_1, n_2\in {\mathbb Z}
\ee
we need 
\bea
F_1(s) 
&\stackrel {\rm def}=&
\sum_{n_1,n_2} \left|\left(\frac{2\pi}{a}\right) \sqrt{n_1^2+n_2^2}\right|^{-s} \nonumber\\
&=&   \left(\frac a{2\pi}\right)^s \sum_{n_1,n_2} (n_1^2+n_2^2)^{-s/2}\nonumber\\
&=&   \left(\frac a{2\pi}\right)^s \frac{1}{\Gamma(s/2)}\int_0^\infty t^{s/2-1}\left\{ \sum_{n_1,n_2\in {\mathbb Z}} 
e^{-t(n_1^2+n_2^2)} -1\right\}dt.\nonumber\\
\eea
The spectrum allows all  positive or negative $n$'s but in defining $F_1$ we must  exclude the case when both $n_1$ and $n_2$ are zero;  hence the $-1$ in the integral.
With the $n_1=n_2=0$  term absent the   Gauss sum is exponentially small at large $t$, but has a small-$t$ asymptotic expansion arising from the Poisson summation identity
\be
\sum_{n=-\infty}^\infty e^{-tn^2}= \sqrt{\frac{\pi}{t}} \sum_{m=-\infty}^{\infty} e^{-\frac 1{4t} (2\pi m)^2}.
\ee
Corrections  to the leading $\sqrt{ \pi/ t}$ are  $\propto e^{-\#/t}$, and so zero from the viewpoint of asymptotic expansions. The poles in $F_1(s)$ arise from the small-t divergences and so for the purposes of extracting the residues we may write 
\bea
F_1(s) &\sim &  \left(\frac a{2\pi}\right)^s \frac{1}{\Gamma(s/2)} \int_0^1 t^{s/2-1} \left\{\frac{\pi}{t}-1\right\} dt+ {\rm holomorphic}.\nonumber\\
\label{EQ:F1-integral}
\eea 
The upper limit ``1" on the integral (\ref{EQ:F1-integral}) can be replaced by any convenient number without altering the residues, so
\bea
F_1(s) 
 &=&  {\pi} \left(\frac{a}{2\pi}\right)^{s} \frac1 {s/2-1} + {\rm holomorphic }
 \nonumber\\
&=& 2{\pi} \left(\frac{a}{2\pi}\right)^2\frac{1}{s-2} + {\rm holomorphic}.
\eea
There appears to be no poles at $s=1,3$.

The Landau-level sum involving 
\be
E(n,k) =  \sqrt{ \left(\frac{2\pi n}{b}\right)^2 + \frac{4\pi |n| k}{a^2}}, \quad n\in {\mathbb N},\quad  \hbox{degeneracy $2|n|$}
\ee
is more complicated. We need the $s=1,3$ residues of 
\bea
F_2(s) &=& \sum_{n=1}^\infty \sum_{k=1}^\infty 2n \left( \sqrt{ \left(\frac{2\pi n}{b}\right)^2 + \frac{4\pi |n| k}{a^2}}\right)^{-s}\nonumber\\
&=& \left(\frac{b}{2\pi }\right)^{s}\sum_{n=1}^\infty \sum_{k=1}^\infty 2n  \left(  n^2 + \left(\frac{  b^2}{\pi a^2}\right) |n| k \right)^{-s/2}\nonumber\\
&=& \left(\frac{b}{2\pi }\right)^{s}\frac{1}{\Gamma(s/2)}\int_0^\infty t^{s/2-1} \sum_{n=1}^\infty \sum_{k=1}^\infty 2n  e^{-t\left(n^2 + \left(\frac{  b^2}{\pi a^2}\right) |n| k\right) }\,dt.\nonumber\\
\eea
Set $\kappa =   b^2/\pi a^2$ and  perform the sum on $k$ to find
\bea
{F_2(s)} &=& \left(\frac{b}{2\pi }\right)^{s}\frac{1}{\Gamma(s/2)}\int_0^\infty t^{s/2-1} \sum_{n=1}^\infty 2n e^{-tn^2} \frac {e^{-\kappa n t}}{1- e^{-\kappa n t}} \,dt\nonumber\\
&=& \left(\frac{b}{2\pi }\right)^{s}\frac{2}{\Gamma(s/2)}\int_0^\infty t^{s/2-1} \sum_{n=1}^\infty  e^{-tn^2} \frac {n}{e^{\kappa n t}-1} \,dt
\label{EQ:F2-sum}
\eea 
 The generating function for Bernoulli numbers gives
\be
\frac {n}{e^{\kappa n t}-1}=  (\kappa t )^{-1}- \frac 12  n + \frac 1 {12}( \kappa t)n ^2 - \frac 1{720} (\kappa t)^3n^4+\ldots,
\ee
and the  Euler-Maclauren formula gives the  small-$t$ asymptotic expansions for the associated Gauss sums as
 \bea
\sum_{n=1}^\infty e^{-tn^2}&\sim& \frac{\sqrt{\pi}}{2 \sqrt t} - \frac 12 ,\nonumber\\
\sum_{n=1}^\infty n e^{-tn^2}&\sim&\frac 1{2t}- \frac 1{12}- \frac{t}{120} - \frac{t^2}{504}+\ldots,\nonumber\\
\sum_{n=1}^\infty n^2 e^{-tn^2}&\sim&\frac {\sqrt{\pi}} {4t^{3/2}},\nonumber\\
\sum_{n=1}^\infty n^4 e^{-tn^2}&\sim &\frac{3\sqrt \pi}{8 t^{5/2}}.
\label{EQ:Euler-Maclaurin}
\eea
All  lines  other than the second in (\ref{EQ:Euler-Maclaurin})  have no higher terms in the asymptotic expansion.   There are corrections,  but they are
$\propto e^{-\#/t}$. 

Inserting the expansions  (\ref{EQ:Euler-Maclaurin}) into (\ref{EQ:F2-sum}) and replacing the  upper limit on the integral by unity gives a sum of poles at integer  $s\le 3$. In particular    we find that ${\rm Res}_{s=1}[F_2]= b^3/24 \pi^2 a^2$  and ${\rm Res}_{s=3}[F_2]= ba^2/2\pi^2$. Inserting these residues into  (\ref{EQ:hitchin-trick}), and adding in   the contribution of the ``topological'' modes from (\ref{EQ:eta-topological}), we end up with
\bea
\eta(0)&=& \left(\frac{b^4}{a^4}\right) \frac 1{\pi^2} \left(\frac 1{48} + \frac 1{192}- \frac 1 {64}\right)+\frac 16 \nonumber\\
&=&\frac 1{96 \pi^2} \left(\frac{b^4}{a^4}\right)+ \frac 16 .
\eea
In the first-line parentheses  $1/48$ is from the residue of $F_2(s)$ at $s=1$, the $1/192$ is from the residue at $s=3$ and the $-1/64$ from the topological modes. The additional 1/6  is also from topological modes, but does not affect the induced charge. 


\begin{thebibliography}{99}
\bibitem{baerI}
Example 3.5 of \cite{baer}. Note that, perhaps confusingly,   these  authors  do not distinguish between  $x,y,z$ and $\xi,\eta,\zeta$.
\end{thebibliography}
 \end{document}